\documentclass[fleqn,usenatbib]{mnras}

\usepackage[T1]{fontenc}

\DeclareRobustCommand{\VAN}[3]{#2}
\let\VANthebibliography\thebibliography
\def\thebibliography{\DeclareRobustCommand{\VAN}[3]{##3}\VANthebibliography}

\usepackage{graphicx}	
\usepackage{amsmath}	
\usepackage{natbib}
\usepackage{amssymb}
\usepackage{enumitem}
\usepackage{amsmath}
\usepackage{iondefs}

\usepackage{newtxtext,newtxmath}

\title[Assessing CGM metallicities]{Using cosmological simulations and synthetic absorption spectra to assess the accuracy of observationally derived CGM metallicities}

\author[\sc Marra {\etal}]
{Rachel Marra,$^{1}$\thanks{E-mail: rmarra@nmsu.edu}
Christopher W. Churchill,$^{1}$
Caitlin Doughty,$^{1}$
Glenn G. Kacprzak,$^{2,3}$
Jane Charlton,$^{4}$
\newauthor Sameer,$^{4}$
Nikole M. Nielsen,$^{2,3}$
Daniel Ceverino$^{5}$
and Sebastian Trujillo-Gomez $^{6}$
\\
$^{1}$Department of Astronomy, New Mexico State University, Las Cruces, NM 88003, USA\\
$^{2}$Centre for Astrophysics and Supercomputing, Swinburne University of Technology, Hawthorn, Victoria 3122, Australia\\
$^{3}$ARC Centre of Excellence for All Sky Astrophysics in 3  Dimensions (ASTRO 3D), Australia\\
$^{4}$ The Pennsylvania State University, State College, PA 16801, USA\\
$^{5}$ Universidad Aut\'onoma de Madrid, Ciudad Universitaria de Cantoblanco, 28049 Madrid, Spain\\
$^{6}$ Institute for Computational Science, University of Zurich\\
}

\date{Accepted XXX. Received YYY; in original form ZZZ}

\pubyear{2021}

\begin{document}
\label{firstpage}
\pagerange{\pageref{firstpage}--\pageref{lastpage}}
\maketitle

\begin{abstract}
We used adaptive mesh refinement hydrodynamic cosmological simulations of a $z=1$ Milky Way-type galaxy and a $z=0$ Dwarf galaxy and generated synthetic quasar absorption-line spectra of their circumgalactic medium (CGM). Our goal is to assess whether standard observational spectroscopic analysis methods accurately reproduce intrinsic column densities, metallicities [Si/H], and hydrogen densities $n_{\tH}$, in simulated absorption-line systems. Without knowledge of the intrinsic simulated properties (blind study), we analysed synthetic COS and HIRES spectra with fixed $S/N=30$ to determine the column densities, metallicity, and $n_{\tH}$, using Voigt profile fitting combined with Markov-Chain Monte-Carlo single-phase {\sc Cloudy} modelling techniques. To quantify the simulated absorbing gas properties, we objectively determined which gas cells along a line of sight (LOS) contribute to detected absorption in the spectra and adopt the unweighted geometric mean of these properties. For this pilot study, we performed this experiment for five LOS in the two simulated galaxies. We found an average agreement between the ``observed'' and intrinsic metallicity overestimated within $0.8\sigma$ or $0.2$ dex for the ``Milky-Way'' and overestimated within $1.4\sigma$ or $0.2$ dex for the Dwarf galaxy. We found that the spectroscopically-derived $n_{\tH}$ are underestimated within $0.8\sigma$ or $0.4$ dex of the intrinsic $n_{\tH}$ for the ``Milky-Way'' and overestimated within $0.3\sigma$ or $0.3$ dex for the Dwarf galaxy. The overall agreement suggests that, for single-phase ionisation modelling of systems where there is substantial spread in gas properties, global metallicity measurements from quasar absorption line studies are capturing the {\it average\/} metallicity and ionisation parameters.  
\end{abstract}


\section{Introduction}
A central goal of astrophysics is to understand how galaxies evolve. As the cycling of baryons is a key process, we can learn a great deal about galaxies if we can understand how baryons cycle in and out of galaxies, in what gas phases they persist, and the how their metal abundances are distributed.  These insights would provide information as to how the baryon cycle drives galaxy evolution and gives rise to the observed universe of stars and galaxies. The interplay between stellar feedback processes in the interstellar medium (ISM) and filamentary accretion from the intergalactic medium (IGM) gives rise to the extended ($> 150$ kpc) metal-enriched circumgalactic medium (CGM).  Acting as the interface between the ISM and IGM, the CGM is a key regulating component of galaxies and is therefore critical to understand as the gatekeeper of the baryon cycle.

The baryon cycle provides a fundamental understanding of galaxies, star formation, and chemical evolution. Theory is used to establish the physical processes by which baryonic matter responds to the full spectrum of gravity-induced dark matter overdensities, thereby explaining many observed global galaxy relations such as the stellar mass-to-halo mass ratio and its cosmic evolution, and the stellar mass--metallicity relation; \citep[e.g.,][]{tremonti04,  behroozi13, gu16, sanchez19}. Significant effort has been directed toward understanding the baryon cycle using hydrodynamic cosmological simulations via the tracking of gas cycles and the modelling of star formation and feedback physics \citep[e.g.,][]{keres05, keres09, oppenheimer08, ceverino09, ceverino14, Trujillo15, Kacprzak16,dave16, nelson19}. \citet{dave11} were able to show that star-forming galaxies develop via a slowly evolving equilibrium balanced by inflows (driven by gravity/mass), wind recycling, star formation rates, and outflows, the latter regulating the fraction of inflow that gets converted into stars. The CGM gas content regulates the competition between IGM inflow and gas ISM consumption by star formation \citep[e.g.,][]{birrer14, finlator17}.

The best approach for observationally probing baryons in, around, and between galaxies is to analyse absorption lines in spectra of background quasars whose sight lines pass near galaxies. The key is to quantify and characterise the dynamics, spatial distributions, metallicities, densities, and temperatures of the gas flowing into, out of, and through galaxies. These quasar absorption lines indirectly provide the aforementioned gas properties; we say ``indirectly'' because chemical-ionisation models of the gas are required to take incomplete information from various absorbing transitions of various ions in order to extract the physical conditions of the gas. Furthermore, a quasar line of sight (LOS) provides a pencil beam 1D probe of the gas as a function of LOS velocity (3D velocity dotted into the LOS).  

Interpretation of such data in terms of the global evolution of galaxies has been guided by hydrodynamic simulations \citep[see][for a review]{TumlinsonReview17}.  However, the vast breadth of our collective understanding is based on the analysis of the spectra by observers, who chart the densities, metallicities, ionisation conditions, and kinematics of the gas.  \citep[e.g.,][]{stocke13, werk14, lehner14, Lehner18, lehner19, Wotta16, Wotta19,  prochaska17, Pointon_2019}.  The important astrophysical conclusions of these works guide our inferences about galaxy evolution and the baryon cycle. As such, continued investigation into the methods is merited.

The problem is that the hydrodynamic simulations clearly show that the analysis methods applied by observers can run completely counter to the intrinsic nature of CGM gas. Observers can be blind to the much richer reality of the physics occurring along the quasar LOS. For example, in simulations, there can be multiple density peaks that overlap in LOS velocity such that their column density contributions in a spectral line are not fully separable; these velocity-aligned yet spatially separated peaks can have different densities, temperatures, and/or metallicities \citep{churchill15, liang18, peeples19}.

This is counter the central assumptions by which observers analyse the data by fitting Voigt profiles (VP) to the absorption line profiles. VP fitting is founded on the assumption of multiple spatially discrete isothermal clouds, each having a unique LOS velocity. VP fits yield the column densities and velocities of the detected ions (such as Mg$^+$, C$^{+3}$, O$^{+5}$, etc.) for each cloud.  The time-honoured method is to use these column densities to constrain chemical-ionisation models, such as {\sc Cloudy} \citep{Ferland17}. The {\sc Cloudy} models provide the critical gas properties, such as hydrogen number density ($n\tH$), metallicity ($Z/Z_{\odot}$), ionisation parameter ($U=n_\gamma/n_{\tH}$), and ionisation conditions ($\log U$); these are key quantities describing the CGM. 

Zoom-in Eulerian cosmological simulations incorporate sophisticated stellar feedback processes into the baryon physics \citep[supernovae, runaway stars, radiation pressure, radiative heating, etc.,][]{ceverino10, ceverino14,  Trujillo15}. Some have experimented with ``forced resolution'' for a more refined investigation into the CGM gas structures and their kinematics \citep[e.g.,][]{Hummels19, peeples19}. This level of realism provides an open road for applying observational quasar absorption line analysis methods to simulated galaxies.  There is now an opportunity for a revolutionary paradigm shift in our ability to interpret quasar absorption line data. 

Though simulations are commonly touted as a powerful tool for providing deeper insights into the interpretation of quasar absorption line data, and many forays in this direction have been undertaken \citep[e.g.,][]{churchill15, liang18, Kacprzak_2019, peeples19, peroux20, marra21, strawn21}, no experiment has been performed as to how the standard spectral analysis plus chemical-ionisation modelling methods employed by observers hold for simulated quasar absorption lines, i.e., do we recover the intrinsic mean physical properties of the {\it absorbing gas\/} using standard VP fitting to the spectra \citep[see][and references therein]{churchill20} followed by {\sc Cloudy} \citep{Ferland17} ionisation modelling constrained by the Voigt profile column densities?

Our goal with this paper is to conduct a double-blind pilot study to directly examine how well the long-standing observational analysis methods of applying single-phase ionisation models to metal-rich quasar absorption line systems capture the intrinsic underlying gas properties.  In particular, we focus on the metallicity and hydrogen density (ionisation parameter) of the absorbing gas systems. By ``double blind'', we mean that we perform this experiment on two parallel tracks: (1) individuals are blindly handed mock absorption line spectra from the simulations following which they perform a full observational analysis of the absorbing gas, and (2) individuals are blindly provided the simulation gas properties that are responsible for the detected absorption lines and perform and analyse the intrinsic simulated gas properties.  We then compare the two parallel analyses.    

In Section \ref{Simulations} we describe the cosmological simulations we use. In Section \ref{Methodology} we discuss the software used to create synthetic spectra from sightlines through the simulations and our pilot study. In Section \ref{Analysis} we explain how we determine metallicity and hydrogen density using the synthetic spectra and observational techniques. In Section \ref{Results} we present our comparison of the metallicity and hydrogen number density values using our different analysis methods. We summarise our findings, discuss them in the context of quasar absorption lines studies, and provide concluding remarks in Section \ref{Discussion}. Throughout we adopt an $H_{0} = 70$~{\kms}~Mpc$^{-1}$, $\Omega_{\tM} = 0.3$, $\Omega_{\Lambda} = 0.7$ cosmology.

\section{Simulations}
\label{Simulations}

\begin{figure*}
\centering
\includegraphics[width=0.95\hsize]{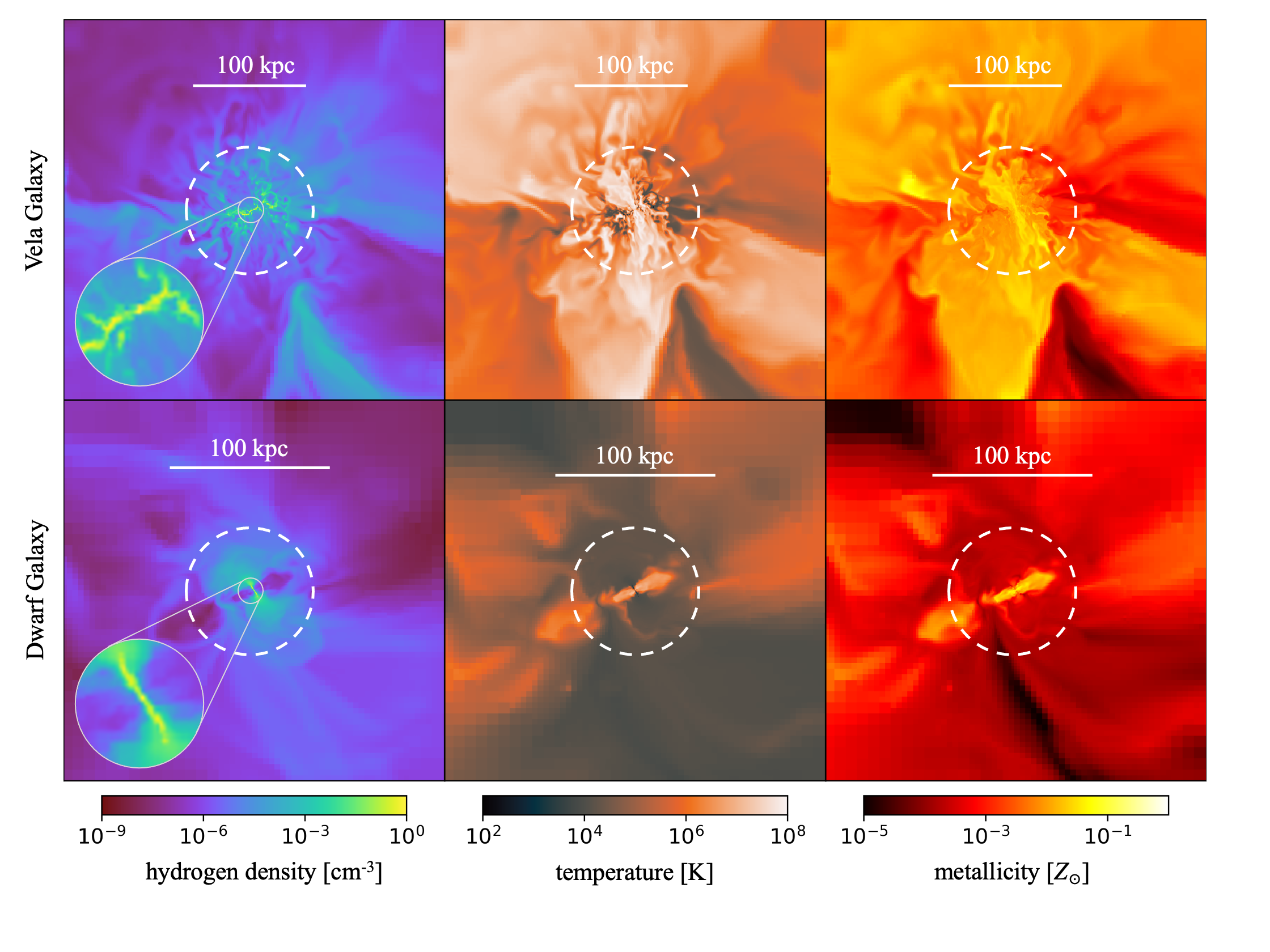}
\vglue -0.3in
\caption{Random slices centred on the simulated galaxies used in this study. (upper row) The $z=1$ galaxy VELA27, which has a virial radius of 112~kpc \citep{ceverino14}. (lower row) The $z=0$ Dwarf galaxy, which has a virial radius of 79~kpc \citep{Trujillo15}.  (left to right) The gas hydrogen number density [cm$^{-3}$], gas temperature [K], and gas phase metallicity [$Z_{\odot}$], respectively. Each slice shows a region extending to $3R_{vir}$; the virial radius is shown as the dashed circle. The insets in the density panels show the inner $0.1R_{vir}$ to provide a detailed look at the structure of the gas disk.}
\label{PrettyPic}
\end{figure*}

In this work we studied two different simulated galaxies. The first is named ``VELA27'', a higher-mass, Milky Way-type galaxy at a redshift of $z = 1$ from the VELA simulations of \citet{ceverino14}. The second is named ``D9m4a'', a low-mass Dwarf galaxy at a redshift of $z = 0$ from the  simulations of \citet{Trujillo15}.

Both galaxies were simulated using the Hydrodynamic Adaptive Refinement Tree code known as {\sc ART} \citep{Kravstov97, Kravstov99_thesis, Kravtsov_2003, ceverino09, Trujillo15}. The {\sc ART} code combines dark matter $\Lambda$CDM cosmological simulations using an N-body {\sc ART} code and Eularian methods to treat hydrodynamics while employing the zoom-in technique of \citet{Klypin2001}.

A galaxy is identified for hydrodynamic simulation in a sphere with a radius of two times the virial radius, generally around 1--2 comoving Mpc. The sphere is traced back to its Lagrangian volume at $z = 60$, and the zoom-in technique \citep[see][]{Klypin2001} is applied to refine the fluctuations down to the chosen resolution limit. We first add gas to the box following the dark-matter distribution using  the universal baryonic fraction. We then re-simulated the whole box, refining resolution in the selected Lagrangian volume around the simulated galaxy. The integration of the gas physics and gravity are performed on the adaptive mesh. The refinement of each cell size is the result of slicing a cell into 8-subcells, with each having half the 1D size and one-eighth the volume.

The two selected simulated galaxies were each created using different stellar feedback recipes, gas heating and cooling subgrid physics, and Eulerian mesh gas cell resolutions.  We emphasise that, in this work, we are not investigating the outcomes of the CGM properties as a function of the subgrid physics employed in the simulations. We aim to assess only how accurately observational spectroscopic absorption line analysis methods recover the probed CGM gas properties. The simulations have slightly different maximum resolutions (one with a maximum resolution of 20~pc, the other 57~pc), which may allow us to obtain some degree of insight into whether gas cell resolution affects this accuracy \citep[however, see][]{Hummels19, peeples19, vandevoort19}. 

\begin{table}
\centering
\caption{Simulated Galaxy Properties \label{tab:galaxy}}
\begin{tabular}{r c c c c c} 
\hline\hline 
 & $\log(M_{vir})$ & $\log(M_{*})$ & $R_{vir}$ & SFR \\[-0.7ex] & [M$_{\odot}$] & [M$_{\odot}$] & [kpc] & [M$_{\odot}$~yr$^{-1}$]\\ [0.5ex] 
\hline 
VELA ($z=1$) &  11.6 & 10.3 & 110 & 0.7 \\
D9m4a ($z=0$) &  10.5 & 7.3 & 79 & 0.002 \\
\hline 
\label{table1}
\end{tabular}
\end{table}

The basic properties of these galaxies are listed in Table~\ref{tab:galaxy}. Both simulated galaxies are described in greater detail in Sections~\ref{vela} and \ref{D9m4a}. These two galaxies were selected with no {\it a priori\/} knowledge of their properties, such as their star formation rates, and stellar or halo masses. In Figure~\ref{PrettyPic}, we show slices through the gas distribution centred on the galaxy for both simulations (VELA top row, Dwarf bottom row) showing gas hydrogen number density (left), temperature (centre), and  metallicity (right). A 100 kpc scale is indicated and the dashed circles show the virial radius. The inset in the density panel highlights the inner 0.1$R_{vir}$.

\subsection{The VELA Galaxy}
\label{vela}

We selected the massive galaxy \#27 from the VELA simulations, which use the RadPre\_LS\_IR feedback model described in \citet{ceverino14}. This model differs from the RadPre runs used in \citet{zolotov15}. In brief, stellar feedback includes thermal and radiative feedback that incorporates the effects of radiation pressure of ionising and infrared photons, local photo-ionisation and photo-heating around young and massive stars. Gas is self-shielded, advects metals, undergoes metallicity-dependent cooling, and is heated by a homogeneous ultraviolet background. The gas can cool to 300~K due to metal and molecular line cooling. Gas flows, shock fronts, and metal disbursement follow self-consistently from this physics. 

The highest spatial resolution of a gas cell at $z=1$ is approximately 20~pc, a regime in which stellar feedback overcomes radiative cooling \citep{ceverino09}. This results in natural galactic outflows \citep{ceverino10, ceverino16} and allows for a combination of cold flow accretion, mergers, and galactic outflows that results in galaxy formation and evolution proceeding on physically-based principles. We note that these simulations have been generally successful at producing observed galaxy properties, such as the distribution of halo masses, stellar mass to halo mass function, mass metallicity relation, Tully-Fisher relation, rotation curves, halo mass to star formation rate, etc.\ \citep{ceverino14}.

The full VELA simulations are 20 Mpc on a side. We extracted a smaller post-production box centred on the target galaxy that is roughly four virial radii ($4 R_{vir}$) in diameter.  The VELA27 galaxy has a maximum cell resolution of $17$~pc, a minimum stellar particle mass of $10^{3}$~M$_{\odot}$, and a dark matter particle mass of $8.3  \!\times\!   10^{4}$~M$_{\odot}$. At $z=1$, the galaxy has $\log(M_{vir}/{\rm M}_{\odot}) = 11.6$,  $\log(M_{*}/{\rm M}_{\odot}) = 10.3$,  $R_{vir} = 110$~kpc, and a star formation rate of 0.7~M$_{\odot}$~yr$^{-1}$.

\subsection{The Dwarf Galaxy}
\label{D9m4a}

The main differences between the VELA simulations and the Dwarf simulations are the star formation and feedback subgrid physics implemented in the simulations, the full details of which can be found in \citet{Trujillo15}.  

In the Dwarf galaxy simulations, star particles are formed in a denser ($n_{\tH} \sim 100$ cm$^{-3}$), colder ($T \sim 100$ K) molecular phase. Furthermore, stars are formed ``deterministically'' based on observations of star forming efficiency in star forming regions in molecular clouds, which is $\sim\! 2$--3\%. Hydrodynamically, the effect of photo-ionisation heating and radiation pressure was included in the feedback implementation of the Dwarf simulation by adding a non-thermal dynamical pressure to the gas surrounding young star particles, based on expectations for HII regions. This combination of star formation and stellar feedback was found to accurately reproduce many of the properties seen in low-mass galaxies at z $\approx$ 0 such as the baryon content, stellar-to-halo mass ratio, star formation history, cold gas fraction galaxy morphology, and rotation curves \citep[see][]{Trujillo15}. 

Similar to the VELA simulations, the Dwarf galaxy simulations have  high-resolution hydrodynamic regions surrounding the galaxies that extend $\sim 1$--2 Mpc in diameter. As with the VELA galaxies, we extracted smaller post-production boxes centred on the target galaxies that are roughly four virial radii ($4 R_{vir}$) in diameter. 

The Dwarf galaxy has a maximum cell resolution of $57$~pc, a minimum stellar particle mass of $100$~M$_{\odot}$, and a dark matter particle mass of $9.4 \!\times\!  10^{4}$~M$_{\odot}$. At $z=0$, the galaxy has $\log(M_{vir}/{\rm M}_{\odot}) = 10.5$,  $\log(M_{*}/{\rm M}_{\odot}) = 7.3$,  $R_{vir} = 79$~kpc, and a star formation rate of 0.002~M$_{\odot}$~yr$^{-1}$.

\section{Methodology}
\label{Methodology}

We conducted a double-blind study to analyse the simulated CGM gas using two common approaches: theoretical and observational. We began by running lines of sight (LOS) through the CGM of the simulated galaxies and generating synthetic absorption line spectra. The absorption lines were then located in the spectra and measured (as described in Section~\ref{Mockspec}). 

The analysis from the theoretical perspective directly examined the properties of the gas cells in the simulated CGM (see Section~\ref{AbsAnal}). The analysis from the observational perspective involved inferring the average gas properties using observer-based spectral analysis methods, including Voigt profile fitting and ionisation modelling (see Section~\ref{cloudy}). Both these analysis perspectives were performed separately of one another by different individuals and then were compared to examine and quantify the degree to which the observer analysis captured the intrinsic average properties of the simulated CGM gas (the theoretical analysis).

\subsection{Synthetic Absorption Line Generation}
\label{Mockspec}

We use the {\sc Mockspec}\footnote{https://github.com/jrvliet/{\sc Mockspec}} pipeline \citep[see][]{churchill15, rachel_thesis} to generate the synthetic quasar spectra and analyse the resulting absorption features. 

Every gas cell in ART has a unique 3D spatial coordinate, physical size ($L_{\rm cell}$), 3D velocity components, temperature ($T$), hydrogen number density ($n_{\tH}$), and metal mass fraction ($x_{\tM}$). We perform a post-processing equilibrium ionisation modelling to determine the ionisation fractions in order to calculate the number densities of all ion stages. We employed the photo+collisional ionisation code {\sc Hartrate} \citep[detailed in][]{cwc14}, which defaults to solar abundance mass fractions for each individual metal up to zinc \citep{Draine, Asplund09}. However, we are capable of adjusting the abundances on an element-by-element basis to match other astrophysically motivated abundance patterns. The {\sc Hartrate} code is well suited for studying the low-density CGM ($\log n_{\tH}/{\rm cm}^{-3}  <-1$) as it gives the best results for optically thin, low density gas.\footnote{Note the total {\HI} column density for all but two of our selected LOS are optically thick, $\log (N_{\tHI}/{\rm cm}^{-2}) > 17.2$. This would suggest that the gas should experience some self-shielding which tends to favour lower ionisation in the shielded regions, which could affect metallicity estimates.  However, the post-processing ionisation corrections are conducted on a cell-by-cell basis, and most cells are optically thin. The code does not account for the cumulative column densities of cells locally embedded in higher density neutral regions.}

Previous studies \citep{cwc1317b, kcn12, churchill15, Kacprzak_2019, marra21} have used {\sc Hartrate}. A quantitative comparison with the industry-standard ionisation code {\sc Cloudy} \citep{Ferland98,Ferland13} shows that across astrophysically applicable densities and temperatures for optically thin gas, the ionisation fractions are in agreement within $\pm 0.05$ dex. In the post-production boxes, all the gas cells are illuminated with the ultraviolet background (UVB) spectrum of \cite{HaardtMadau2005} to obtain the equilibrium solution. {\sc Hartrate} records the electron density, ionisation and recombination rate coefficients, ionisation fractions, and number densities for each gas cell and for all ions from hydrogen through zinc.

\begin{table}
\begin{center}
\caption{LOS Data}
\begin{tabular}{c c c c c c} 
\hline\hline 
Sim. & LOS & $D$ [kpc] & $\log (N_{\tHI}/{\rm cm}^{-2})$ & $D/R_{vir}$ \\[0.5ex] 
\hline 
VELA27 & 0022 &  7.3 & 18.52 & 0.06 \\
VELA27 & 0025 &  8.1 & 18.79 & 0.07 \\
VELA27 & 0027 &  8.8 & 17.85 & 0.08 \\
VELA27 & 0055 &  18.6 & 16.67 & 0.17 \\
VELA27 & 0081 &  30.5 & 14.20 & 0.28 \\
\hline
D9m4a & 0034 &  3.9 & 17.47 & 0.05  \\
D9m4a & 0061 &  6.9 & 17.77 & 0.09 \\
D9m4a & 0072 &  8.4 & 17.77 & 0.11  \\
D9m4a & 0091 &  10.3 & 18.13 & 0.13  \\
D9m4a & 0137 &  15.3 & 17.41 & 0.19  \\
\hline 
\label{table2}
\end{tabular}
\end{center}
\vglue -0.3in
\end{table}

We use {\sc Mockspec} to generate a user-specified number of "quasar" LOS that are distributed through a simulated galaxy. Relative to the target simulated galaxy, each LOS is defined by the position angle on the plane of the sky (in the range of $0^{\circ} \leq \phi \leq 360^{\circ}$), the impact parameter, and the sky-projected inclination of the galaxy. The plane of the sky is defined as the plane through the centre of mass of the galaxy that is perpendicular to the LOS.  The details of the methodology to produce synthetic spectra from quasar sightlines and details on how the profiles are synthesised are outlined in \citet{churchill15}.  The absorption features in the synthetic spectra are objectively detected using methods detailed in \cite{churchill00} as originally developed by \cite{schneider93}.

As this is a pilot study, we limited our analysis to ten LOS systems, using five LOS from the VELA27 simulated galaxy and five LOS from the simulated Dwarf galaxy. The LOS ID number, impact parameter, $D$, the {\HI} column density, and normalised impact parameter, $D/R_{vir}$, are listed in Table \ref{table2}. The VELA27 galaxy has an inclination angle of $90^{\circ}$ and the Dwarf galaxy has an inclination angle of $0^{\circ}$.  This selection was arbitrary. All LOS had randomly selected position angles. For each simulated galaxy, the five LOS were selected by using simple selection criteria. We avoided LOS that were damped {\Lya} absorbers (DLAs). This is because the ionisation corrections for DLAs are often assumed to be zero or negligible, and we wanted to test the more common sub-Lyman limit and Lyman-limit systems. 

Preliminary examination of the gas cells along the LOS provided estimates for the {\HI} column  density.  Our selection criteria resulted in primarily Lyman limit systems. In order to have robust constraints from the metal lines, we selected systems for which {\CII}, {\CIII}, {\CIV}, {\SiII}, {\SiIII}, {\SiIV}, and {\MgII} absorption lines were detected with rest-frame equivalent widths of $0.1$~{\AA} or greater. We did not require the same detection criteria for the {\OVI} and {\FeII} metal lines.  From the VELA galaxy, we included two sub-Lyman limit systems, LOS0055 and LOS0081.  The latter has only metal absorption lines for {\CIII}, {\CIV}, and {\OVI}. Our reasoning is that we wanted to examine at least two lower {\HI} column density systems. 

\begin{figure*}
\centering
\includegraphics[width=0.98\hsize]{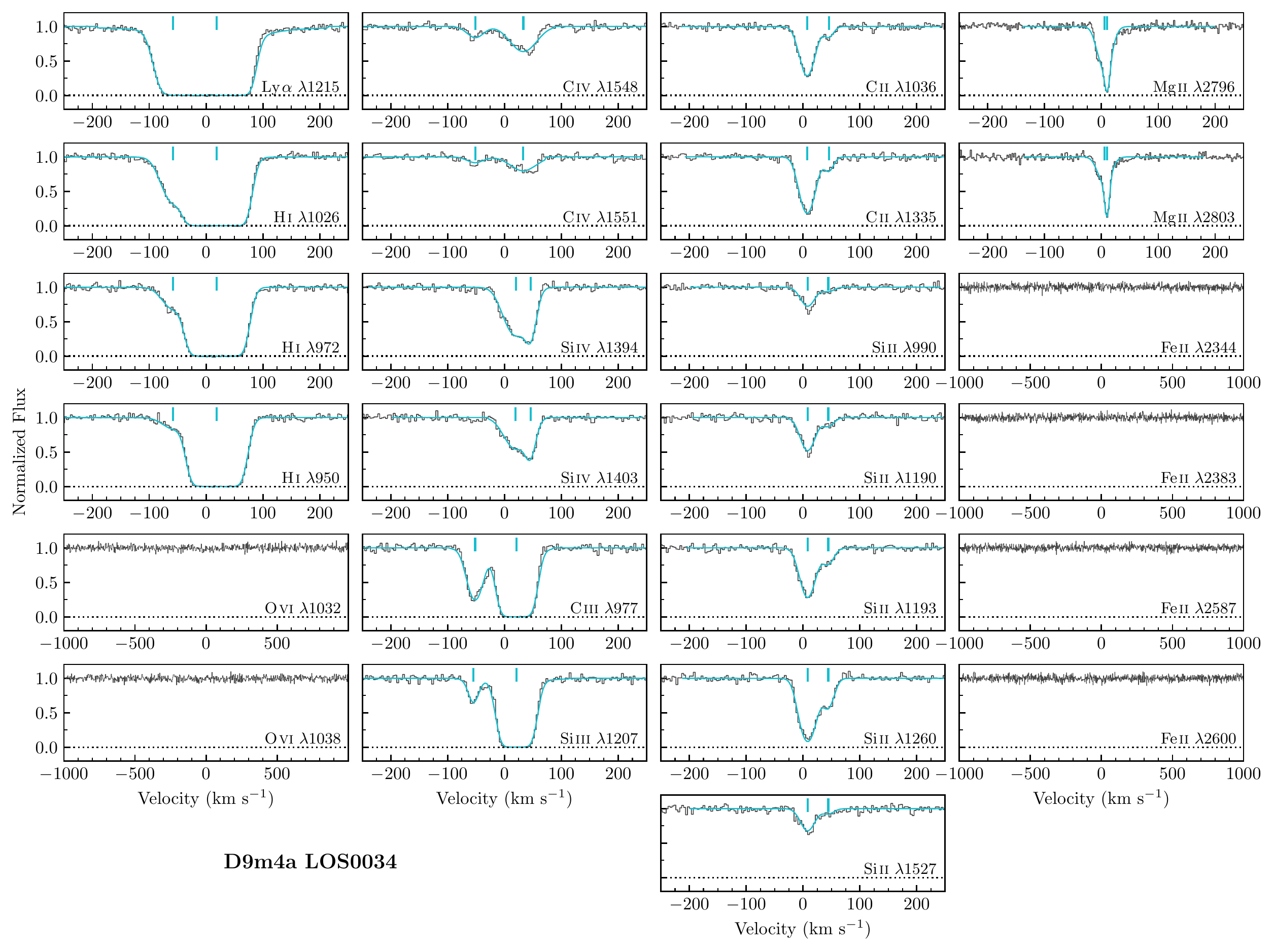}
\caption{The synthetic absorption line spectra and Voigt Profile (VP) fitted spectra (cyan curves) for LOS0034 of the $z=0$ Dwarf galaxy D9m4a. Ticks above the spectra provide the number of VP components and the velocity centre of each. The VP fitting is used to measure the total column density of each ion, in this case {\HI}, {\CIV}, {\CIII}, {\CII}, {\SiIV}, {\SiIII}, {\SiII}, and {\MgII}, from the absorption lines of commonly observed transitions. Upper limits ($3\sigma$) on column densities are determined from the equivalent width detection threshold assuming a Doppler parameter of $b=8$~{\kms}. The absorbing gas cells selected from these spectra are shown in Figure~\ref{los0034}. The synthetic absorption line spectra and VP fitted spectra for the other LOS can be found online in the supplementary material.}
\label{vpfit}
\end{figure*}

\begin{figure*}
\centering
\includegraphics[width=0.95\hsize]{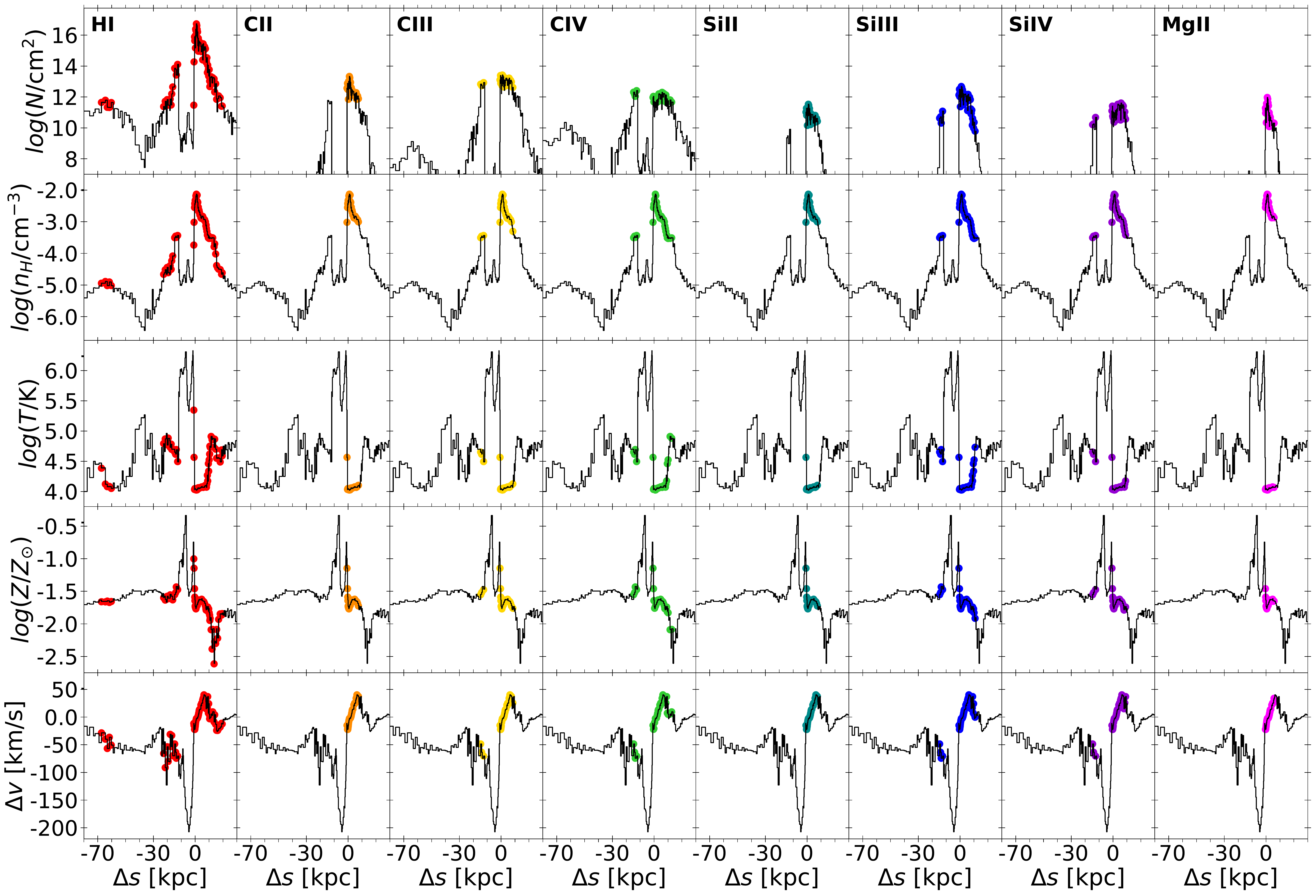}
\caption{The physical conditions of the gas cells intercepted by LOS0034 for the $z=0$ Dwarf galaxy as a function of line of sight position $\Delta S$ (in kpc), where $\Delta S=0$~kpc is the plane of the sky intersecting the centre of mass of the simulated galaxy. (across, left to right) The properties for the ions {\HI}, {\CII}, {\CIII}, {\CIV}, {\SiII}, {\SiIII}, {\SiIV}, and {\MgII} are shown. (top to bottom) The gas cell column density ($N_{\rm ion}$ [cm$^{-2}$]), the hydrogen number density ($n_{\tH}$ [cm$^{-3}$]), the temperature ($T$ [K]), the metallicity (Z[$Z_{\odot}$]), and the line of sight velocity ($\Delta v$ [{\kms}]), respectively. The column densities are computed from $N_{\rm ion} = f_{\rm ion} n_{\rm elem} \Delta L$, where $f_{\rm ion}$ is the ionisation fraction of the chemical element of density $n_{\rm elem}$, and $\Delta L$ is the actual path length of the line of sight through the cell (not the cell wall length). Absorbing cells for each ion are marked using coloured dots; these cells are selected from the synthetic absorption line spectra as described in Section \ref{DefAbsCells}. This figure for the other LOS can be found online in the supplementary material.}
\label{los0034}
\end{figure*}

As an illustrative example, the synthetic spectra of Dwarf LOS0034 are shown in Figure~\ref{vpfit}. We show commonly observed transitions for {\HI} (including Ly$\alpha$, Ly$\beta$, Ly$\gamma$, and Ly$\delta$), {\CIV}, {\CIII}, {\CII}, {\SiIV}, {\SiIII}, {\SiII}, {\MgII}, {\OVI} and {\FeII}.  The black lines show the spectra generated with {\sc Mockspec}, the cyan lines show the Voigt Profile fitted spectra (see Section~\ref{vpfitting}), and the cyan tick-marks show the velocity centre of each VP component. For transitions with rest-frame wavelengths in the far UV (FUV), we adopt a spectral resolution of $R=20,000$ with eight pixels per resolution element to emulate the G130M and G160M gratings of the Cosmic Origins Spectrograph \citep[COS,][]{green-cos} on board the {\it Hubble Space Telescope} ({\it HST\/}). However, instead of using the actual COS line spread function, we adopted a simple Gaussian function.  For transitions in the near UV (NUV), we adopted $R=45,000$ with three pixels per resolution element to emulate the HIRES \citep{vogt-hires} and UVES \citep{dekker-uves} spectrographs on the Keck and Very Large Telescopes, respectively.  All synthetic spectra have a signal-to-noise-ratio $S/N=30$. These spectral characteristics yield a $3\sigma$ detection threshold sensitivity of $W_r \simeq 0.05$~{\AA} for the ``COS'' spectra and  $W_r \simeq 0.02$~{\AA} for the ``HIRES'' spectra.

\begin{figure*}
\centering
\includegraphics[width=0.94\hsize]{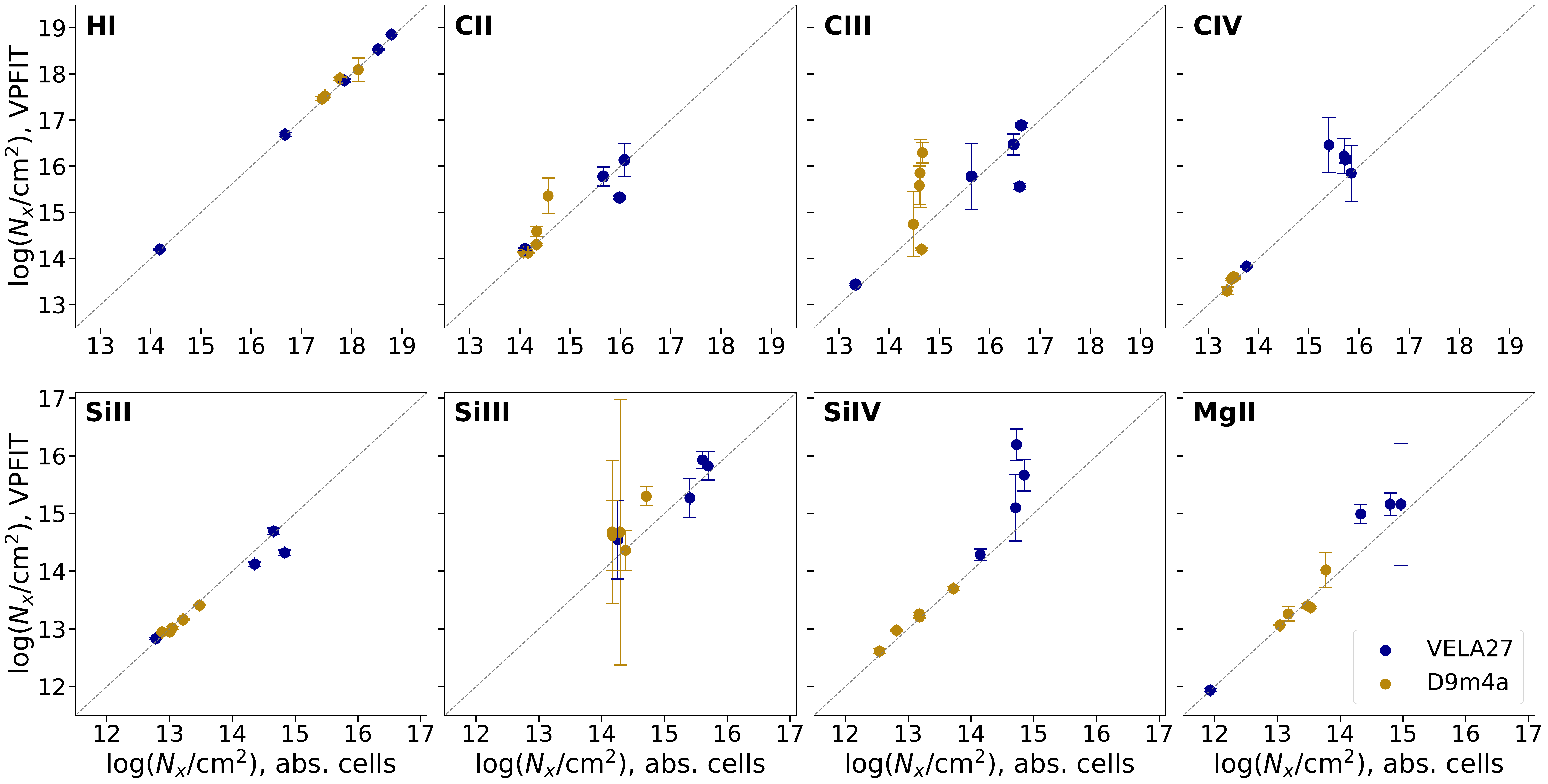}
\caption{Comparison of the VP fitted total column densities for each ion ``X'' and the sum of the absorbing cell column densities for that ion. The vertical error bars are the $1\sigma$ uncertainties in the VP values; there are no errors in the sums for the cells. Blue points are the five sight lines for the $z=1$ VELA  galaxy and gold points are the five sight lines for the $z=0$ Dwarf galaxy.}
\label{logN}
\end{figure*}

\section{Analysis}
\label{Analysis}
 
 \subsection{Theoretical Approach: Defining Absorbing Gas Cells}
\label{DefAbsCells}

Once synthetic spectra are generated, {\sc Mockspec} determines which gas cells along a given LOS contribute to the absorption features objectively detected to the $5\sigma$ significance level in the synthetic spectra.  There are several criteria that {\sc Mockspec} uses to make this determination. For gas cells pierced by the LOS, the ionic column density of each cell is calculated using the path length of the LOS through the cell. Cells with ion column density $N_{i,x} < 10^9$~cm$^{-2}$ are assumed to not contribute to detectable absorption. {\sc Mockspec} identifies which of the remaining gas cells give rise to the observed absorption by determining which gas cells account for 95\% of the equivalent width of the absorption features. The cells are sorted by decreasing column density and are removed until the resulting (noiseless) spectrum created by the remaining cells has an equivalent width equal to 95\% of the original spectrum's equivalent width.  Remaining gas cells (typically around 10-30) are considered ``absorbing cells'' as they are the cells along the LOS that contribute 95\% of the measured equivalent width. The process is repeated for each ion, yielding the absorbing cells for each ion.  

In Figure~\ref{los0034}, we show the physical conditions of the gas cells intercepted by LOS0034 for the $z=0$ Dwarf galaxy as a function of line of sight position $\Delta S$ (in kpc), where $\Delta S=0$~kpc is the plane of the sky intersecting the centre of mass of the simulated galaxy. Each column from left to right corresponds to eight of the ions we studied, {\HI}, {\CII}, {\CIII}, {\CIV}, {\SiII}, {\SiIII}, {\SiIV}, and {\MgII}. We highlight the absorbing cells for each ion as the coloured points.  

It is clear that the majority of the gas cells intersected by LOS0034 do not contribute to the detected absorption features in the synthetic spectra.  This clearly illustrates the need for comparing only the absorbing cells (instead of all intersected gas cells) for understanding the connection between spectral absorption lines and the physical properties of the gas giving rise to absorption.   To this point, we use only the absorbing gas cells for our analysis. For example, the majority of the metal-line absorption is spatially concentrated at $\Delta S \simeq 5$--20~kpc and this gas arises in a contiguous grouping of cells with $T \!\simeq\! 10,000$~K in a $n_{\tH}$ density enhancement in the range $-2 \geq \log n_{\tH}/{\rm cm}^{-3} \geq -3$. There is a velocity shear with $\Delta v \simeq 40$~{\kms} over which the $n_{\tH}$ density decreases as $\Delta v$ is increasing across the shear. 

Quite importantly, the metal-line absorption we study does not arise in the highest metallicity gas, which is associated with a grouping of lower density, $\log n_{\tH}/{\rm cm}^{-3} \simeq -4.5$, hotter $T\simeq 10^6$~K gas cells. The high velocity ($\Delta v \simeq -50$ to $-200$~{\kms}) of these lower density, hotter, metal-enriched cells may constitute a galaxy outflow (note it is blue shifted and on the near side of the galaxy). The rapid velocity inversion adjacent to the absorbing cells is suggestive of a shock front that has collided into slightly redshifted, higher density, cooler, lower metallicity absorbing gas. Interestingly, the intermediate and higher ions, {\CIII}, {\CIV}, {\SiIII}, and {\SiIV}, and the neutral hydrogen {\HI}, also have some absorbing cells on the opposite side of the putative shock front (at $\Delta S \leq -10$~kpc).

\subsection{Theoretical Approach: Mean Properties of Absorbing Cells}
\label{AbsAnal}

Our goal is to determine a single mean value for the metallicity and for the number density of hydrogen in a LOS. Following the methods of \citet{Pointon_2019}, we use [Si/H] to quantify metallicity as opposed to $Z/Z_{\odot}$. For details on how [Si/H] is computed for each absorbing cell, see the Appendix.

To calculate a single averaged value of [Si/H] and $n_{\tH}$ for a given LOS, we calculate the geometric mean values of [Si/H] and $n_{\tH}$ from the absorbing cells, 
\begin{equation}
\langle X \rangle = 
\left\{ \prod _{i=1}^{n}x_{i}\right\} ^{1/n}={\sqrt[{n}]{x_{1}x_{2}\cdots x_{n}}} \, , 
\label{geomean}
\end{equation}
where $n$ is the number of cells. The geometric mean is employed as it is appropriate when the values span several orders of magnitude, whereas the arithmetic mean is dominated by the largest values. The standard deviation in the geometric mean is
\begin{equation}
\sigma _{\langle X \rangle}
= \exp \left\{ \frac{1}{n} \displaystyle \sum _{i=1}^{n}\left[\log\left(\frac{x_{i}}{\langle X \rangle}\right)\right]^{2} \right\}
 \, . 
\label{gstd}
\end{equation}

Pursuant our goal to compare the geometric mean values to the values determined using observational methods (described in Section~\ref{cloudy}), we include all unique absorbing cells that give rise to {\HI}, {\CII}, {\CIII}, {\CIV}, {\SiII}, {\SiIII}, {\SiIV}, {\MgII}, and {\FeII} absorption in  Eq.~\ref{geomean}. That is, we count cells only once even if they gave rise to absorption in multiple metal lines.

\subsection{Observational Approach: The Absorption Lines}
\label{vpfitting}

The synthetic spectra were used for the absorption line analysis in the same fashion as would occur for an observer. The observational analysis was done without any information about the underlying properties of the simulated galaxies. Using the synthetic spectra, the column densities of each ion were obtained using Voigt profile (VP) fitting. We used version 12.2 of {\sc VPfit} \citep{vpfit}. The initial guesses for {\sc VPfit} were generated manually using its interactive mode feature, and its iterative fitting procedure derived a best-fit model upon achieving a sufficiently good $\chi^2$. We obtain VP component column densities, Doppler $b$ parameters, redshifts, and the $1\sigma$ uncertainty on each parameter. For a given ion, ``X'', the column densities of the VP components are summed to obtain the system total column density $\log N_{\tion}$. 

We ``averaged'' the gas conditions across the full profiles by using $\log N_{\tion}$ to constrain the ionisation models.  We thus adopted a fitting philosophy focused on minimising the $\chi^2$ for each ion.  Though we enforced that transitions of the same ion had identical column densities, Doppler $b$ parameters, and velocity components, we did not ``tie together'' the velocity centres and/or Doppler $b$ parameters of different ions of similar ionisation potentials, a technique commonly applied for VP fitting that focuses on decomposing the absorption profiles kinematically. In Figure~\ref{vpfit}, we show the {\sc VPfit} absorption line models superimposed on the synthetic spectra for LOS0034. 

To check that the {\sc VPfit} column densities accurately reflect the column densities of the absorbing gas cells, we compared $\log N_{\tion}$ (from {\sc VPfit}, see Figure~\ref{vpfit}) and the sum of the column densities of the absorbing cells (see Figure~\ref{los0034}). In Figure~\ref{logN}, we present {\sc VPfit} column densities versus the summed absorbing cell column densities for {\HI}, {\CIV}, {\CIII}, {\CII}, {\SiIV}, {\SiIII}, {\SiII} and {\MgII}.  The gold points correspond to the five LOS through the Dwarf galaxy and the blue points correspond to the five LOS through the VELA galaxy. 

It is reassuring that the {\sc VPfit} column densities generally match the summed absorbing cell column densities.  To a high degree, the values agree within the $1\sigma$ uncertainties in the {\sc VPfit} column densities. This would suggest that the {\sc VP} modelling of synthetic absorption line spectra accurately reflect the column densities of the absorbing cells. As the column density is effectively the product of the number density and LOS path length, this provides some confidence that the synthetic spectra and {\sc VP} modelling reflect the underlying density distribution of the absorbing cells.

\subsection{Observational Approach: The Ionisation modelling}
\label{cloudy}

To estimate the mean [Si/H] and $n_{\tH}$ of the absorbing cells from synthetic absorption, we used the Markov-Chain Monte-Carlo (MCMC) technique described in \citet{Pointon_2019}, originally developed by \citet{crighton15}. Following \citet{Pointon_2019}, we generate {\sc Cloudy} grids using version 13.05 \citep{Ferland13}, spanning a range of hydrogen number densities and metallicities that are plausible for the CGM, with a {\HI} column density range selected to bound the $N_{\tHI}$ value derived from {\sc VPfit} results. We adopted the UV background ionising spectrum of \citet{HaardtMadau2005}, calculated for the average redshift of the line fits returned by {\sc VPfit}. Note that this is the same UV background used to compute the ionisation fractions of hydrogen and the metals in the gas cells of the simulations. The {\sc Cloudy} grids are used with the {\sc \emph{emcee}} package \citep{foreman-mackey13} to perform MCMC sampling and create the posterior distributions for [Si/H], $\log N_{\tHI}$, $\log n_{\tH}$, and $\log U$ based on the metal and hydrogen column densities derived from {\sc VPfit}. 

The {\HI} column density is assumed to have a Gaussian prior centred on the measured $N_{\tHI}$, and all other parameters are assumed to have flat priors spanning the entire range included in the {\sc Cloudy} grids. The MCMC process is run using 200 walkers for 200 steps to perform the burn-in, and an additional 200 steps for the MCMC sampling.  Although the {\OVI} lines were VP fitted when detected, following \citet{werk14, Wotta16, Wotta19, Pointon_2019}, the O$^{+5}$ ion was not included in the MCMC modelling. As such, our resulting MCMC/{\sc Cloudy} [Si/H], $\log N_{\tHI}$, $\log n_{\tH}$, and $\log U$ output results should be viewed as a single-phase average of multi-ion absorption systems for which the {\OVI} absorption phase is not weighted in the solution.   We remind the reader that the analysis was done blind of any details about the underlying properties of the simulated galaxies and used only the synthetic spectra provided.

\begin{figure}
\centering
\includegraphics[width=0.99\hsize]{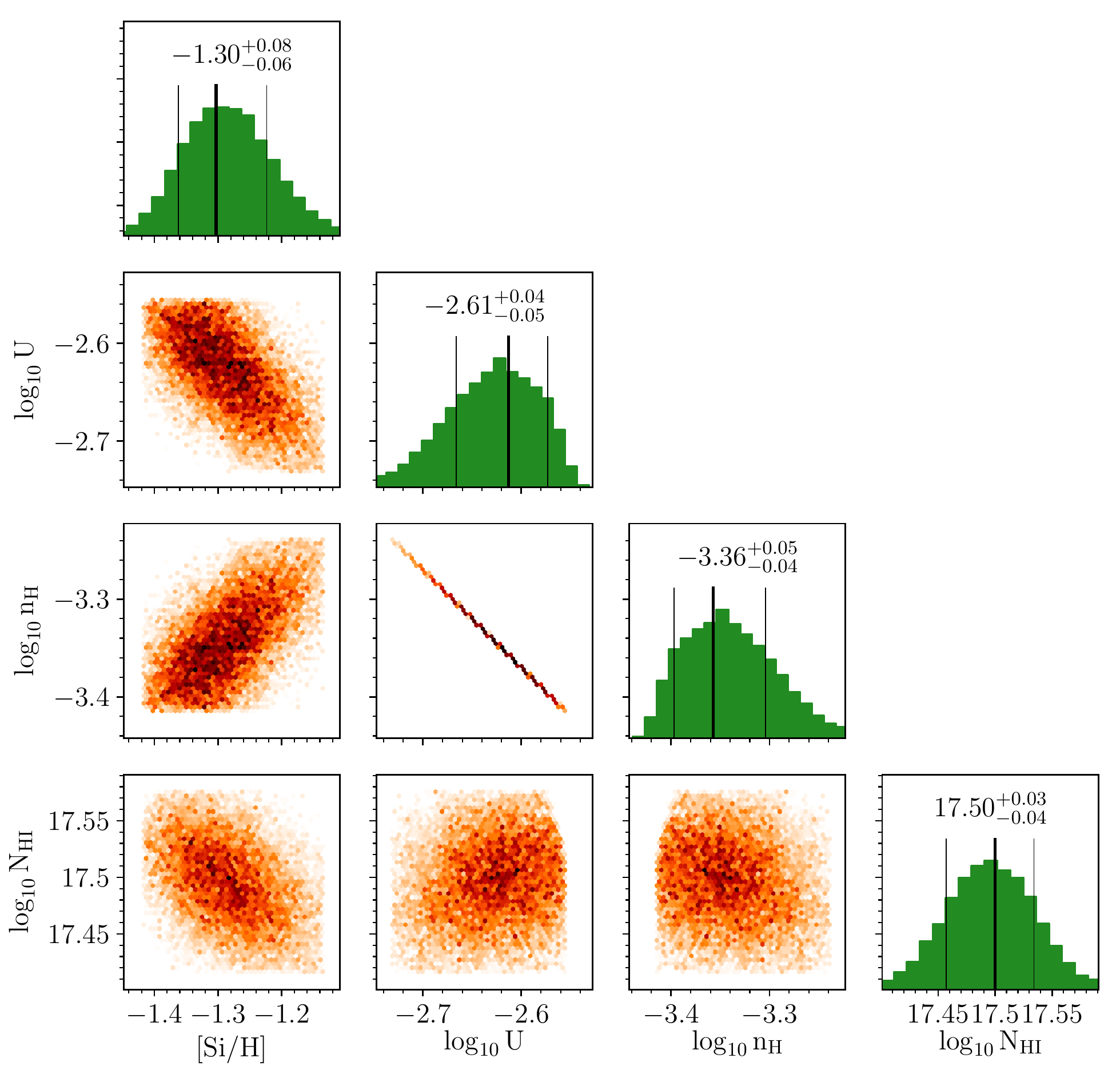}
\caption{Results of the Markov Chain Monte Carlo \citep[see][]{crighton15} chemical ionisation modelling using {\sc Cloudy} \citep{Ferland17} applied to the absorption derived column densities of LOS0034 of the $z=0$ Dwarf galaxy. Shown are the silicon abundance [Si/H], the ionisation parameter, $U$, the hydrogen number density, $n_{\tH}$, and the neutral hydrogen column density, $N_{\tHI}$. The orange points provide the 2D density distributions of viable models; the green distributions provide the 1D distributions.  The means and $1\sigma$ uncertainties are the adopted values.}
\label{Corner34}
\end{figure}

An example of the MCMC/{\sc Cloudy} results is shown in Figure~\ref{Corner34} for Dwarf LOS0034. The panels with the green histograms show the posterior distributions of [Si/H], $\log N_{\tHI}$, $\log n_{\tH}$, and $\log U$, along with vertical black lines depicting the most likely value (thick centre line) and $1\sigma$ values (two bracketing thin lines). These MCMC/{\sc Cloudy} values are to be compared with the geometric mean values of the absorbing gas cells.

\section{Results}
\label{Results}
 
\begin{figure*}
\centering
\includegraphics[width=0.8\hsize]{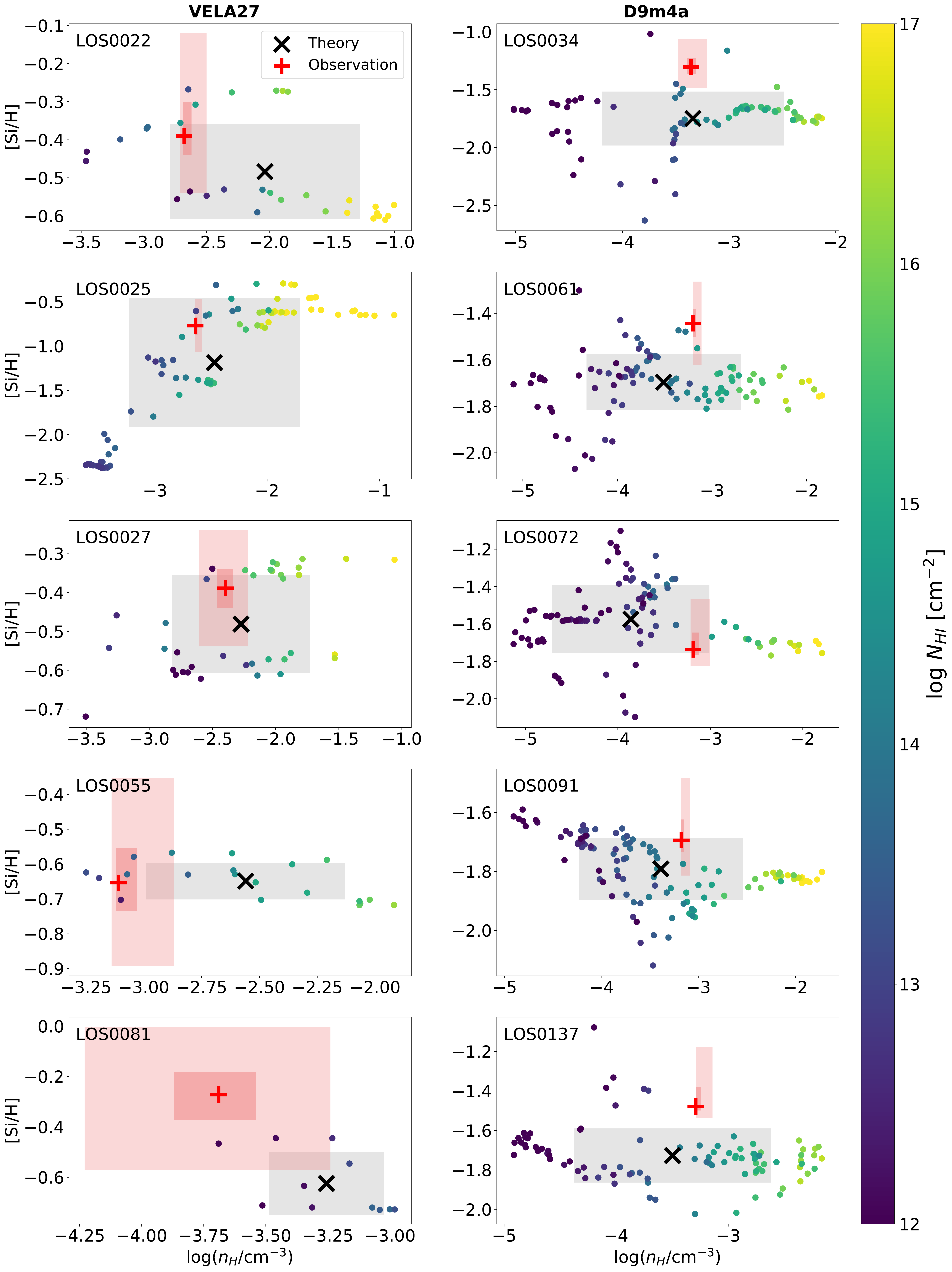}
\caption{Metallicity, [Si/H], versus hydrogen number density, $n_{\tH}$ for (left panels) the five sight lines for the $z=1$ VELA  galaxy and (right panels) the five sight lines for the $z=0$ Dwarf galaxy. Data points are the absorbing cells, colour coded by their {\HI} column density. The black ``X" corresponds to the geometric mean [Si/H] and $n_{\tH}$ of the cells. The light grey shading shows the $1\sigma$ uncertainty in the geometric means. The MCMC/{\sc Cloudy} results, based on {\sc VPfit} $\log N_{\tion}$ measured from the synthetic spectra, are shown as the red ``{\bf +}'', including the $1\sigma$ (dark pink shading) and $3\sigma$ uncertainties (light pink shading). }
\label{scatter}
\end{figure*}

For the ten LOS examined in this study, we show the comparison between the mean [Si/H] and $n_{\tH}$ properties obtained using the absorbing cells and the absorption line analysis (MCMC/{\sc Cloudy} modelling) in Figure~\ref{scatter}.  Each panel shows [Si/H] versus $n_{\tH}$ for a single LOS, with LOS0022, 0025, 0027, 0055, and 0081 from the VELA simulations in the left-hand panels and LOS0034, 0061, 0072, 0091, and 0137 from the Dwarf simulations in the right-hand panels. Each data point corresponds to an absorbing gas cell (those that contribute to detectable absorption in the synthetic spectra). Absorbing cells are coloured by their {\HI} column density as indicated in the colour bar ranging from $12 \leq \log (N_{\tHI}/{\rm cm}^{-2}) \leq 17$. The black ``X" gives the geometric mean value of [Si/H] and $n_{\tH}$ of the absorbing cells, calculated using Equation~\ref{geomean}. The light grey shaded ``box'' region corresponds to the $1\sigma$ uncertainties of the geometric mean, calculated using Equation~\ref{gstd}.

Regarding the absorbing cell properties, there are some systematic differences between those selected from the VELA and Dwarf galaxies.  Compared to the Dwarf galaxy, in the VELA galaxy there are generally fewer absorbing cells. This is especially true for LOS0055 and 0081, which have the lowest $N_{\tHI}$ absorption column densities. In general, there are $\sim\! 20$--30 absorbing cells for the LOS through the VELA galaxy. Conversely, for each LOS of the Dwarf galaxy, there are $\sim\! 50$ absorbing cells. Considering the mean cell size of absorbing gas cells in the different galaxies, the VELA27 galaxy has a mean absorbing cell size of 226~pc while the Dwarf galaxy has a larger mean absorbing cell size of 460~pc. The slightly larger average for the Dwarf galaxy is because the absorbing cell sizes are roughly equally distributed between the two cell sizes 218~pc and 436~pc, whereas $\simeq\! 95$\% of the VELA galaxy absorbing cells correspond to the 218~pc cell size.   

Additionally, though the range of $n_{\tH}$ of the absorbing cells are fairly similar between the VELA and the Dwarf galaxies ($-3.5 \leq \log n_{\tH}/{\rm cm}^{-3} \leq -2.0$), the [Si/H] values are roughly an order of magnitude higher for the absorbing gas cells from the VELA galaxy as compared to the Dwarf galaxy.  For the VELA galaxy, the range is roughly $-0.8 \leq \hbox{[Si/H]} \leq -0.2$, whereas the range for the Dwarf galaxy is roughly $-2 \leq \hbox{[Si/H]} \leq -1$. This difference is likely because the VELA galaxy is significantly more massive than the Dwarf galaxy and the greater stellar mass has yielded more mass in metals in the VELA CGM. Also, the VELA galaxy is at a redshift of $z \approx 1.0$ while the Dwarf is at $z \approx 0.0$, and there are generally more galactic outflows at higher redshift. 

LOS0025 from the VELA galaxy has a number of absorbing cells comparable to those found for the Dwarf galaxies and also a broader distribution of [Si/H] values than the other LOS from the VELA galaxy. Interestingly, for the LOS0022, 0025, and 0027 for the VELA galaxy, the distribution of [Si/H] values also exhibit bimodalities in the [Si/H] values with respect to the geometric mean [Si/H] of the absorbing cells. Examination of the global distribution of the absorbing cell [Si/H] for five LOS through the Dwarf galaxy show that it is a Gaussian distribution about the geometric mean value.  

In Figure~\ref{scatter}, the measured $n_{\tH}$ and [Si/H] derived from the MCMC/{\sc Cloudy} modelling based on the $\log N_{\tion}$ determined from {\sc VP} modelling of synthetic absorption lines in the synthetic spectra are marked with a red ``{\bf +}". The $1\sigma$ uncertainties are the darker pink shaded ``box'' regions centred on the red ``{\bf +}" and the $3\sigma$ uncertainties are the lighter pink shaded regions.  Before comparing the MCMC/{\sc Cloudy} results to the properties of the absorbing gas cells in the simulations, we remind the reader that the MCMC/{\sc Cloudy} modelling was performed blind of any information about the underlying properties of the simulated galaxies. Only the synthetic spectra were made available for analysis in the identical fashion as would occur for an observer. After this ``observer'' followed the analysis steps described in Sections~\ref{vpfitting} and \ref{cloudy} the resulting MCMC/{\sc Cloudy} results were then made available for comparison with the theoretical approach using the mean absorbing gas properties.

In all cases, the $3\sigma$ uncertainty box from the MCMC/{\sc Cloudy} modelling overlaps with a subset of the absorbing gas cells. However, LOS0081 only has two absorbing cells overlapping with the $3\sigma$ uncertainty box. There are only eleven absorbing cells for this LOS, which is significantly fewer absorbing cells than any of the other LOS. Note that this is the lowest $N_{\tHI}$ system in our study, with $\log N_{\tHI}/{\rm cm}^{-2} = 14.2$. The discrepancy between the MCMC/{\sc Cloudy} analysis and the cell values could result because this is the only LOS that has spectroscopically detected metal-absorption \textit{only} in {\CIV} and {\CIII} while all other LOS have metal absorption from  {\CII}, {\CIII}, {\CIV}, {\SiII}, {\SiIII}, {\SiIV}, {\MgII}, and occasional {\FeII}. It is also possible that the discrepancy was a result of the single phase assumption breaking down as there are no low ions in this system.

In order to quantify the level of overlap between the observer MCMC/{\sc Cloudy} method and the theoretical geometric mean values from the absorbing cells, we computed a ``$z$-score'' for both the [Si/H] and $n_{\tH}$ for each LOS.  For the ``observed'' quantity $x$ with geometric ``theoretical'' mean $\langle X \rangle$, the $z$-score is simply $z$-score$ =(x - \langle X \rangle )/\sigma$, where $\sigma = [ \sigma^2(x) + \sigma^2(\langle X\rangle )]^{1/2}$, and where $\sigma(x)$ and $\sigma(\langle X\rangle )$  are the one-sided standard deviations between $x$ and $\langle X \rangle$, respectively. We list the $z$-score for each LOS and the mean score for the VELA and Dwarf galaxy in Table~\ref{table3}. The $z$-score accounts for both the uncertainty in the ``observer'' value and the dispersion in the absorbing cell property and is interpreted as the number of $\sigma$ the two values $x$ and $\langle X \rangle$ depart from one another; we thus quote the $z$-score in $\sigma$ units.

\begin{figure*}
\centering
\includegraphics[width=0.92\hsize]{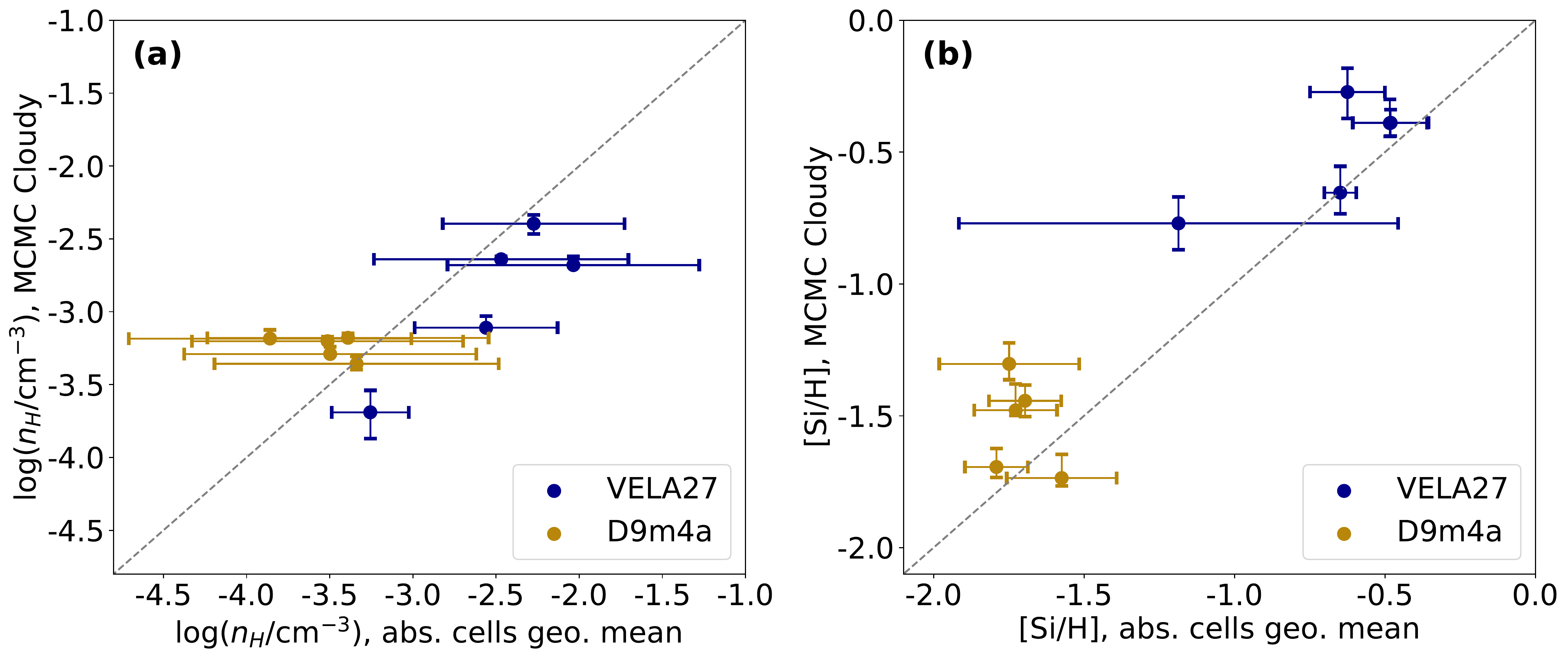}
\caption{Comparison of (a) $n_{\tH}$ and (b) [Si/H] from the MCMC/{\sc Cloudy} analysis of the VP fitted synthetic spectra and the geometric mean of the values for the absorbing gas cells, with $1\sigma$ error bars. Blue points are the LOS for the $z=1$ VELA  galaxy and gold points are the LOS for the $z=0$ Dwarf galaxy.}
\label{nH_SiH}
\end{figure*}

Examining the VELA galaxy results (left panels of Figure~\ref{scatter}), we found that, with the exception of LOS0081, [Si/H] is recovered well within the $1\sigma$ uncertainties of the MCMC/{\sc Cloudy} result. By recovered, we mean that the MCMC/{\sc Cloudy} results are consistent with the geometric mean determined from the absorbing gas cell. On average, we found that the LOS in the VELA galaxy are within $\simeq\!0.8\sigma$ for [Si/H]. However, for all VELA LOS, the MCMC/{\sc Cloudy} modelling systematically underestimated the mean $n_{\tH}$ by $\simeq\!0.8\sigma$. As the ionisation parameter is defined as $U=n_\gamma(z)/n_{\tH}$, where $n_\gamma(z)$ is a fixed photon number density at redshift $z$ for the choice of UV ionising background, we see that the ionisation parameter would be correspondingly overestimated in proportion to $\sigma(U)/U = \sigma(n_{\tH})/n_{\tH}$.

Examining the Dwarf galaxy results (right panels of Figure~\ref{scatter}), we found that for four of the LOS, the MCMC/{\sc Cloudy} metallicities are overestimated by $0.8$ to $1.9\sigma$, while the final LOS is underestimated by $0.8\sigma$. On average the $n_{\tH}$ is recovered within $\simeq\! 0.34\sigma$, with all of the five LOS having their MCMC/{\sc Cloudy} $n_{\tH}$ overestimated relative to the geometric mean of the absorbing cells.

\begin{table}
\begin{center}
\caption{z-score values}
\begin{tabular}{c c c c } 
\hline\hline 
Sim. & LOS & $n_{\tH}$ z-score & [Si/H] z-score \\ [0.5ex] 
\hline 
VELA27 & 0022 & $0.85$ &  $-0.61$  \\
VELA27 & 0025 & $0.22$ &  $-0.56$ \\
VELA27 & 0027 & $0.22$ &  $-0.68$ \\
VELA27 & 0055 & $1.26$ &  $-0.05$ \\
VELA27 & 0081 & $1.48$ &  $-2.21$ \\
\hline
\textbf{VELA27} & \textbf{mean} & $\mathbf{0.81}$ &  $\mathbf{-0.82}$ \\
\hline\hline
D9m4a & 0034 & $-0.02$ &  $-1.86$ \\
D9m4a & 0061 & $-0.38$ &  $-1.89$ \\
D9m4a & 0072 & $-0.79$ &  $0.79$ \\
D9m4a & 0091 & $-0.25$ &  $-0.87$ \\
D9m4a & 0137 & $-0.24$ &  $-1.79$ \\
\hline
\textbf{D9m4a} & \textbf{mean} & $\mathbf{-0.34}$ &  $\mathbf{-1.44}$ \\
\hline 
\label{table3}
\end{tabular}
\end{center}
\vglue -0.3in
\end{table}

A direct comparison of the MCMC/{\sc Cloudy} derived [Si/H] and $n_{\tH}$ values with the theoretical approach values (the mean absorbing cell values) is presented in Figure~\ref{nH_SiH}. Blue points correspond to the LOS through the VELA galaxy, whereas gold points correspond to LOS through the Dwarf galaxy.  In Figure~\ref{nH_SiH}(a), we see that the derived MCMC/{\sc Cloudy} values of $n_{\tH}$ are systematically lower by an average of $0.8\sigma$ or $0.4$ dex than the geometric mean value of the absorbing gas cells for the VELA galaxy and systematically higher by an average of $0.3\sigma$ or $0.3$ dex for the Dwarf galaxy. In Figure~\ref{nH_SiH}(b), we see better overall agreement within $0.2$ dex between the MCMC/{\sc Cloudy} derived [Si/H] and the geometric mean value of the absorbing cells. The best agreement is found for the VELA galaxy, on average within $0.81\sigma$ or $0.19$ dex too high. In the case of the Dwarf galaxy, on average the MCMC/{\sc Cloudy} values are about $\simeq\! 1.44\sigma$ or $0.24$ dex too high.  
 
From the complex scatter distributions of absorbing cell gas properties seen in Figure~\ref{scatter}, it is clear that the geometric mean is a crude single-valued description of a complex distribution of gas properties.  Yet, the MCMC/{\sc Cloudy} modelling formalism employed by most observational studies enforces a single-valued derived quantity for [Si/H] and $n_{\tH}$ for comparison with a complex distribution of absorbing gas properties. As such, this forces an open question as to what is the most appropriate mean value of the simulated absorbing gas to employ for comparison with the MCMC/{\sc Cloudy} results.

In addition to the non-weighted geometric means that we present in Figure~\ref{scatter}, we also explored an $N_{\tHI}$ weighted geometric mean, motivated by the notion that cells with high {\HI} column would contribute more strongly to the absorption lines. We did not investigate a cell mass weighted mean, as suggested by \citet{liang18}, because the LOS path length through a given cell may comprise only a small fraction of the cell wall length. Instead, we account for cell wall length when computing the cell column density, $N_{\tHI}$, which reflects the actual product of $n_{\tH}L$, where L is the path length pierced through the cell. As such, the $N_{\tHI}$ weighting is a proxy for mass weighting. However, we found that weighting the geometric mean with $N_{\tHI}$ resulted in a larger disagreement between the MCMC/{\sc Cloudy} derived $n_{\tH}$ and the weighted geometric mean values. Depending on the LOS, the weighted geometric mean $n_{\tH}$ was an additional $0.5$ to $1$ dex higher when using $N_{\tHI}$ weighting. Comparatively, the [Si/H] values were not significantly different between the weighted and the unweighted geometric mean.  

Based on the $n_{\tH}$ results, we decided to use the unweighted geometric mean to compare with the MCMC/{\sc Cloudy} results, which we believe better reflects the distribution of absorbing cells in terms of their relative contribution to metal line absorption. For example, inspection of Figure~\ref{los0034} for LOS0034 shows that the highest metallicity cells correspond to cells with low {\HI} column densities. This might provide a partial physical explanation for why the MCMC/{\sc Cloudy} $n_{\tH}$ value is found near the cluster of lower $N_{\tHI}$ absorbing cells (see Figure~\ref{scatter}). 

Also apparent from a study of Figure~\ref{los0034} for LOS0034
is that, for the higher ions {\CIII}, {\CIV}, {\SiIII}, and {\SiIV}, there is a small spatially isolated cluster of absorbing cells at $\Delta S \simeq -15$~kpc and a second grouping $\Delta S \simeq +10$~kpc with lower $n_{\tH}$, roughly $\log (n_{\tH}/{\rm cm}^{-3}) \simeq -3.5$. The $n_{\tH}$ densities of these absorption cell groupings are systematically below the density range of the absorbing cells of the low ions {\CII}, {\SiII}, and {\MgII}, which are in the range $-3 \leq \log (n_{\tH}/{\rm cm}^{-3}) \leq -2$.  This is suggestive that the absorption systems is characterised by multiple phases.

\section{Summary and Conclusions}
\label{Discussion}

Our aim with this pilot study of 10 LOS was to develop insights into the efficacy of the techniques applied to quasar absorption lines for determining the physical conditions of the absorbing gas in the CGM. In particular, we examined the common approach of VP fitting the absorption line profiles followed by chemical-ionisation modelling to estimate the metallicity and density (ionisation parameter) assuming a single gas phase.  

We produced synthetic absorption line spectra \citep{churchill15, rachel_thesis} through high-resolution hydrodynamic cosmological simulations \citep{ceverino14, Trujillo15}. We used objective absorption line finding and measuring software \citep{schneider93, weakI} to identify and quantify the absorption lines in the synthetic spectra.  We then manually VP fitted the absorption lines using {\sc VPfit} \citep{vpfit} to determine the column densities of the {\HI} and metal ions.  These column densities were then used to constrain the MCMC/{\sc Cloudy} models \citep{crighton15, Ferland13} to determine the metallicity ([Si/H]), hydrogen number density ($n_{\tH}$), ionisation parameter ($U$), and a model estimate of the neutral hydrogen column density ($N_{\tHI}$), and their uncertainties. Similar methods have been applied for decades \citep[e.g.][]{bergeron86, steidel90, stocke13, werk14, prochaska17, lehner19, Pointon_2019}. 

We created synthetic spectra with the characteristics the COS G130M and G160M gratings on board {\it HST} and the HIRES and UVES instruments with $S/N=30$, corresponding to a $3\sigma$ equivalent width detection threshold of $\simeq 0.05$~{\AA} (COS) and $\simeq 0.02$~{\AA} (HIRES/UVES). Thus, our findings are comparable to observational studies such as \citet{stocke13}, \citet{werk14}, \citet{Pointon_2019}, and \citet{lehner19}. The results of our synthetic observations and analysis were compared to the properties of the gas in the simulations giving rise to the absorption. This is a key point, as we objectively identify the absorbing cells pierced by the LOS that contribute to the {\it detected\/} absorption profiles \citep{churchill15, rachel_thesis}. To the best of our knowledge this is a unique approach to studying the properties of absorbing gas in simulations. As observed absorption lines must represent only the gas giving rise to the absorption actually detected in the spectra, it seems the theoretical studies of simulations should universally adopt this physical fact. This also implies that the gas being studied in absorption is dictated by the detection threshold, or signal-to-noise ratio and resolution, of the spectra.  

To compare the ``observational'' analysis to the absorbing gas cell properties, we employed the unweighted geometric mean to provide a single-valued characterisation of the diverse absorbing gas cell properties (see Eq.~\ref{geomean}). We focused on the metallicity [Si/H] and the hydrogen number density $n_{\tH}$, as these are key quantities that characterise the CGM.   Our main results are presented in Figures~\ref{logN}, \ref{scatter}, and \ref{nH_SiH} as described in Section~\ref{Results}. Our findings can be summarised as follows: 

\begin{itemize}
    \item The summed VP column density for each ion determined from VP fitting the synthetic spectra are generally statistically consistent with the summed column densities of the absorption selected gas cells. This suggests a meaningful correspondence between the properties measured from the synthetic spectra and the selected absorbing cells in the simulations. 
    \item The $n_{\tH}$ derived from the MCMC/{\sc Cloudy} modelling tends to be lower than the unweighted geometric mean of the absorbing cells for the VELA galaxy, and is consistent, on average, to $\simeq\! 0.81\sigma$ or $0.4$ dex. The MCMC/{\sc Cloudy} modelling tends to be higher and does significantly better for the Dwarf galaxy, yielding results consistent within $\simeq\! 0.34\sigma$ or $0.3$ dex.  
    \item The [Si/H] metallicity derived from MCMC/{\sc Cloudy} modelling tends to be somewhat higher than the unweighted geometric mean of the absorbing cells. Typically, for the VELA galaxy, the MCMC/{\sc Cloudy} value is consistent with the geometric mean value to the $\simeq\! -0.82\sigma$ level or $0.2$ dex. The MCMC/{\sc Cloudy} modelling did not match quite as well for the Dwarf galaxy, which on average had results consistent with the geometric mean value to the $\simeq\! -1.4\sigma$ level or $0.2$ dex.
\end{itemize}

For this study, we have assumed a single-phase of gas for the MCMC/{\sc Cloudy} modelling. This means we did not attempt to capture the more detailed distributions of the absorbing gas cell properties as clearly illustrated in Figure~\ref{scatter} (also see Figure~\ref{los0034}).  In addition to not examining the multi-phase gas properties, we followed the modelling philosophy of \citet{werk14, Wotta16, Wotta19, Pointon_2019} and omitted {\OVI} absorption from our analysis. The high ionisation potential of O$^{+5}$ typically places the ion in a lower density, diffuse phase than the absorption from the lower ionisation species \citep[e.g.,][]{bergeron06, cwc_charlton, ding03, muzahid15, rosenwasser18}, or in a collisionally ionized phase \citep[e.g.,][]{haislmaier21, Sameer21}. Further, simulations suggest that {\HI} and {\OVI} absorption, even if aligned in velocity, do not originate from the same physical gas structures \citep[e.g.,][]{oppenheimer09, churchill15}. 

Our approach differs from a similar experiment conducted by \citet{liang18} in which they compare properties (density, temperature, and metallicity) derived using a Bayesian approach with intrinsic gas properties found by {\HI}-mass weighting {\it all\/} gas cells along a LOS. Again, we note that a key to our experiment is that we do not compare all of the gas cells along a LOS, but only those gas cells that significantly contribute to absorption in the synthetic spectra. While \citet{liang18} results are generally consistent with ours, in that both of our methods yield reasonably good agreement between the derived properties and the average of the intrinsic gas properties, they conclude that such agreement would not be achieved under the assumption of a single-phase ionisation model. However, as shown in Figure~\ref{nH_SiH}, our methodology, which is uniquely designed to closely emulate observational techniques, essentially recovers the average intrinsic metallicity of the gas (though with an $\simeq\!0.8\sigma$ or $0.4$ dex underprediction of the gas density for the VELA galaxy and $\simeq\!0.3\sigma$ or $0.3$ dex overprediction for the Dwarf galaxy). 

Perhaps it is not unexpected that our experiment captures the average metallicity of the absorbing gas cells.  As found by \citet[][see their Figure~14]{Sameer21}, a direct comparison between single-phase modelling and multi-phase modelling of complex metal-rich absorption line systems suggests that the single-phase modelling yields the average metallicity of the multi-phase gas properties.  

Interestingly, as can bee seen in Figure~\ref{scatter}, LOS0022, 0025, and 0027 through the VELA galaxy exhibit a metallicity bimodality.  LOS0022 has a grouping of absorbing cells with $\hbox{[Si/H]} \simeq -0.6$ and $\simeq\! -0.3$, whereas LOS0025 has groupings at $\simeq\! -0.5$ and $\simeq\! -1.5$ and LOS0027 has groupings at $\simeq\! -0.6$ and $\simeq\! -0.35$.  \citet{lehner13, Lehner18} and \citet{Wotta16, Wotta19} have reported a bimodal metallicity distribution for $z<1$ partial Lyman Limit systems. As our results suggest that the modelling recovers the average metallicity of a complex distribution of metallicities along the LOS \citep[as also corroborated by][]{Sameer21}, we would conclude that the observed bimodality of lower and higher metallicity systems does not rule out the presence of high metallicity pockets of gas in systems with low average metallicity (and vice versa).  

Indeed, not all metallicity distributions are found to be bimodal.
\citet{prochaska17} found a high metallicity unimodal distribution in $z \simeq 0.2$ Lyman limit systems. Considering these opposing results, an important question to think about is what astrophysics drives such bimodalities or whether the observed bimodality is an artefact of how the metallicity is determined from the data. Our results suggest that the observed metallicity bimodality is not an artefact of the analysis methods used to determine metallicity, but is indicative of truly different {\it averages\/} in these systems. 

What we must understand is that the single-phase modelling does not capture the intrinsic distribution of metallicities in either the ``low-'' and ``high-metallicity'' systems, which may have substantial overlap. An excellent discussion of the short-comings of single-phase modelling are covered by \citet{haislmaier21}. We expect that a multiphase modelling approach, such as those developed by e.g., \citet{Zahedy19,Zahedy21}, \citet{haislmaier21}, or \citet{Sameer21}, applied to LOS0022, 0025, and 0027 through the VELA galaxy should potentially capture the metallicity bimodality in those absorption systems and even the different mean densities of those gas cell groupings.  For example, as evident from the distribution of absorbing cells in Figure~\ref{scatter},  LOS0025 might be best described as a two-phase system with the first phase characterised by $\log n_{\tH}/{\rm cm}^{-3} \simeq -3$ and $\hbox{[Si/H]} \simeq -1.5$ and the second by  $\log n_{\tH}/{\rm cm}^{-3} \simeq -2$ and $\hbox{[Si/H]} \simeq -0.5$.  Even then, chemical-ionisation modelling of absorption lines can, in all practicality, provide only averages of underlying distributions. The distributions of absorbing cell metallicities and hydrogen densities in the simulations bear out that this must be a fundamental truth of observed absorption line systems.

Furthermore, all types of metal-enriched gas structures are not captured in the absorption lines adopted for this study, which by design are typical of the vast majority of quasar absorption line studies.  As illustrated in Figure~\ref{los0034}, along LOS0034 of the Dwarf galaxy, there is a $T \simeq 10^6$~K blueshifted outflow at $\Delta S \simeq 0$ to $-10$ kpc with a peak metallicity of $\log Z/Z_{\odot} = -0.5$ and $\log n_{\tH}/{\rm cm}^{-3} \simeq -4$ and $\log N_{\tHI}/{\rm cm}^{-2} \simeq 10$. This is gas that potentially would give rise to {\NeVIII}~$\lambda\lambda 770,780$ absorption \citep[e.g.,][]{savage05, haislmaier21}, but note that {\it none\/} of the detected {\HI} absorption along this LOS would be associated with this gas phase. This problem of how to ``partition'' the {\HI} absorption between modelled phases presents a tremendous challenge for multi-phase modelling. In our opinion, this example also serves to illustrate why studies that compare the results of observational absorption line analysis methods to the intrinsic properties of the gas in simulations must examine only those gas cells that actually contribute to the detected absorption lines in the synthetic spectra.

As this is a pilot study, we must temper our inferences and conclusions accounting for the limited sample size of 10 LOS. The pilot sample of 10 LOS is a compromise between having a sample that could potentially capture and inform us of possible variations, scatter, and/or systematic errors, while remaining manageable. The observer methodology of {\sc VPfit} plus MCMC/{\sc Cloudy} modelling is human-intensive. Ideally, a fully automated approach would be ideal in order to substantially increase the sample size for which the differences in the densities and metallicities between the observer and simulation methods could be expressed as well-sampled posterior distributions. We are currently working on automating both the {\sc VPfit} and MCMC/{\sc Cloudy} modelling to allow for a future study with a large sample of diverse LOS.

There is also the selection criteria to consider. In selecting the 10 LOS, we prioritised the {\HI} column density to ensure we included Lyman-limit and sub-Lyman limit systems; we then preferentially selected LOS that exhibiting metal-line absorption from a range of ions.  Other than those criteria, we did not base our selection on any other physical properties, such as impact parameter, for example. We only looked at LOS from two simulated galaxies, a massive VELA galaxy and a Dwarf galaxy.  While there are different sub-grid stellar/ISM feedback physics implemented in these two simulations, the hydrodynamic solvers governing the gas physics in the CGM (far from cells in which stellar particles are spatially coincident) are identical for both simulated galaxies. 

It is important to note that our two simulations are at different redshifts; the VELA galaxy at $z = 1.0$, whereas the Dwarf galaxy is at $z = 0.0$. As such, the absorbing gas cells of the two galaxies are subject to different UVB spectral energy distributions. Also of consideration is the hydrodynamic resolution of the simulations. The resolution where the LOS pierce the absorbing gas cells is lower than the maximum resolution of the simulations.  As mentioned in Section~\ref{Results}, for the VELA galaxy we found an mean absorbing cells size of 226 pc while the Dwarf galaxy has a larger mean absorbing cell size of 460 pc. This is counter intuitive when considering our $n_{\tH}$ results, as the VELA galaxy which has the better resolution around the LOS gives less consistent results for $n_{\tH}$ while the Dwarf galaxy has a worse resolution around the LOS and shows more consistent results for this quantity. 

The equilibrium temperature of each gas cell with a given $n_{\hbox{\tiny H}}$ is determined from a cooling curve look-up table for a fixed ``cloud'' thickness of 1~kpc, and, as we have pointed out, the mean resolution (cell sizes) of the absorbing gas cells differ by a factor of about two between the VELA and Dwarf galaxies. The equilibrium temperature strongly effects the ionisation fractions, which effect the metallicity estimates. It would be ideal to control for these variables by employing LOS from both galaxies at the equivalent redshifts and mean resolutions. However, it is computationally expensive to run massive galaxies to $z\sim 0$ with high resolution; as a consequence, the VELA galaxies were run only down to $z=1$. 

Our sample size for this pilot study has shown systematically different results for the LOS in the two different galaxies. In future studies, for which we plan a more in depth study employing thousands of LOS in different simulated galaxies, it will be interesting to see if this statistical difference for the two galaxies hold, or if it is an artefact of the small sample size.

Overall MCMC/{\sc Cloudy} modelling was able to reproduce the geometric mean [Si/H] of the absorbing gas cells within the $0.8\sigma$ level of consistency for the VELA galaxy and $1.9\sigma$ level of consistency for the Dwarf galaxy. We concluded that single-phase MCMC/{\sc Cloudy} models based on VP fit parameters from absorption lines effectively captures the mean metallicity of the gas that is giving rise to absorption. If our results are to be taken at face value, this would suggest that global metallicity measurements from quasar absorption line studies are capturing the {\it average\/} metallicity properties across cosmic time and astrophysical environment.  

We found that the single-phase MCMC/{\sc Cloudy} modelling yielded $n_{\tH}$ values that are systematically too low on average by $0.4$ dex for the VELA galaxy and too low on average by $0.3$ dex compared to the geometric mean $n_{\tH}$ values of absorbing cells.  Though there is roughly a factor of two difference in the sizes of the highest resolution gas cells between these two simulations, we cannot be certain that resolution is affecting the systematic offset. The VELA simulations have higher resolution than the Dwarf simulations, yet the systematic offset is more severe in the VELA simulations. One additional difference between the two simulations is that the number of absorbing cells in a given VELA LOS was smaller than in a given Dwarf galaxy LOS (with the exception of VELA LOS0025). However, there is no trend or correlation between the number of absorbing cells and the offset in the $n_{\tH}$ values. 

In the future we plan a statistical analysis of tens of thousands of LOS using the same methodology used for this pilot study. Our goal is to provide a wholesale statistical quantification of how effectively observational techniques are reproducing physical properties of the gas. We also plan to conduct a pilot study using the multi-phase modelling methods developed by \citet{Sameer21}, including the {\OVI} absorption.

\section*{Acknowledgements}

We thank the referee for helpful comments that improved this manuscript. RM holds an American Fellowship from AAUW. This material is based upon work supported by the National Science Foundation under Grant No.\ 1517816 issued to CWC and JCC. GGK and NMN acknowledge the support of the Australian Research Council through a Discovery Project DP170103470. Parts of this research were supported by the Australian Research Council Centre of Excellence for All Sky Astrophysics in 3 Dimensions (ASTRO 3D), through project number CE170100013. DC is a Ramon-Cajal fellow. DC is supported by the Ministerio de Ciencia, Innovaci\'{o}n y Universidades (MICIU/FEDER) under research grant PGC2018-094975-C21. The VELA galaxy simulation were performed  at NASA Advanced Supercomputing (NAS) at NASA Ames Research Center.

\section*{Data Availability}
The data utilised in this paper will be shared on reasonable request
to the corresponding author.

\bibliographystyle{mnras}
\bibliography{refs} 

\newpage
\appendix
\section{Calculating [Si/H] in a gas cell}
In this Appendix, we describe how we determine the value [Si/H] in a gas cell. We start with the solar mass fractions, 
$(x_{\tH})_\odot = 0.7381$,
$(x_{\tHe})_\odot = 0.2485$,
$(x_{\tM})_\odot = 0.0134$.
The silicon and hydrogen atomic mass units are $A_{\tH} = 1.00784$ and $A_{\tSi} = 28.0855$, respectively. The solar ratio for silicon to hydrogen is $ \log ( n_{\tSi}/n_{\tH} ) _\odot = -4.49$. 
The desired quantity is
\begin{equation}
\left[ {\rm Si/H} \right] = 
\log \left( \frac{n_{\tSi}}{n_{\tH}} \right)
-
\log \left( \frac{n_{\tSi}}{n_{\tH}} \right)_\odot 
 \, .
\label{eq:SiH}
\end{equation}
The relationship between number density and mass fraction
\begin{equation}
\frac{n_{\tSi}}{n_{\tH}}
= \frac{x_{\tSi}/A_{\tSi}}{x_{\tH}/A_{\tH}}  
\qquad {\rm and} \quad 
\left( \frac{n_{\tSi}}{n_{\tH}} \right)_\odot \!\!
= \frac{(x_{\tSi})_\odot/A_{\tSi}}{(x_{\tH})_\odot/A_{\tH}}  
\, .
\label{eq:ntox}
\end{equation}
For the VELA galaxy, which had metallicities typically in the range $\log Z/Z_{\odot} > -1 $, we assumed a solar abundance pattern. For the Dwarf galaxy, which had metallicities typically in the range $\log Z/Z_{\odot} < -1 $, we enhanced $\alpha$-group elements $+0.5$~dex, since one typically sees enhanced $[\alpha/{\rm Fe}]$ for $[{\rm Fe/H}] < 1$   \citep[e.g.,][]{Lauroesch96, Weinberg19}. 

We applied the simple scaling $x_{\tSi}/(x_{\tSi})_\odot
= Z_{\rm cell}/(x_{\tM})_\odot$, where $Z_{\rm cell} = Z_{\tI} + Z_{\tII}$, $Z_{\tI}$ is the mass fraction of metals from Type Ia supernovae yields and $Z_{\tII}$ is the mass fraction of metals from Type II supernovae yields. Taking the ratio of $(n_{\tSi}/n_{\tH})/(n_{\tSi}/n_{\tH})_\odot$ and rearranging, we obtain
\begin{equation}
\frac{x_{\tSi}}{(x_{\tSi})_\odot}
= \frac{Z_{\rm cell}}{(x_{\tM})_\odot} 
= \frac{n_{\tSi}/n_{\tH}}{(n_{\tSi}/n_{\tH})_\odot} \frac{x_{\tH}}{(x_{\tH})_\odot}
\, ,
\end{equation}

which simplifies to 
\begin{equation}
\frac{n_{\tSi}/n_{\tH}}{(n_{\tSi}/n_{\tH})_\odot} 
=
\frac{(x_{\tH})_\odot}{x_{\tH}} \frac{Z_{\rm cell}}{(x_{\tM})_\odot} 
= 
\frac{(Z_{\rm cell}/x_{\tH})}{{[(x_{\tM})_\odot/(x_{\tH})_\odot}]}
\, .
\end{equation}
Taking the log of both sides, we have
\begin{equation}
    {\rm [Si/H] } = \log \left( \frac{Z_{\rm cell}}{x_{\tH}} \right)
    - \log \left( \frac{(x_{\tM})_\odot}{(x_{\tH})_\odot} \right)
   \, ,
\end{equation}
or, 
\begin{equation}
    {\rm [Si/H] }
    = \log \left( \frac{Z_{\rm cell}}{x_{\tH}} \right) + 1.741 \, .
\end{equation}
The hydrogen mass fraction for the cell is obtained from $x_{\tH} + x_{\tHe} + Z_{\rm cell} = 1$, which can be written $x_{\tH} = (1-Z_{\rm cell})/(1+r)$, where $r = x_{\tHe}/x_{\tH}$, giving
\begin{equation}
    {\rm [Si/H] }
    = \log \left( \frac{Z_{\rm cell}(1+r)}{1-Z_{\rm cell}} \right) + 1.741 \, .
\end{equation}
We adopt $r=0.335$. 


\bsp	
\label{lastpage}
\end{document}